\documentclass[aps,prl,twocolumn, superscriptaddress, showpacs]{revtex4-1}
\usepackage{hyperref}
\usepackage{amsmath}
\usepackage{amssymb,amsfonts}
\usepackage{graphicx}
\usepackage{makecell}
\usepackage{blkarray}
\usepackage{mathtools}
\usepackage{relsize}
\usepackage{lmodern}
\usepackage{slantsc}
\usepackage{scalefnt}
\usepackage{subfigure}
\usepackage{dsfont}
\usepackage{xspace} 
\usepackage{boxedminipage}
\usepackage{ifpdf}
\usepackage{pgffor}
\usepackage{xifthen}
\usepackage{tensor}
\usepackage{array} 
\newcolumntype{C}{>{$}c<{$}} 
\usepackage{tikz}
\usetikzlibrary{shapes.geometric}
\usepackage{color}
\usetikzlibrary{arrows,matrix,calc,scopes,decorations.markings}
\allowdisplaybreaks[1]

\newcommand{\be}{\begin{equation}}
\newcommand{\ee}{\end{equation}}
\newcommand{\bse}{\begin{subequations}}
\newcommand{\ese}{\end{subequations}}
\newcommand{\ket}[1]{|{#1}\rangle}

\newcommand{\Z}{\mathbb{Z}}

\newcommand{\bpm}{\begin{pmatrix}}
\newcommand{\epm}{\end{pmatrix}}
\newcommand{\bmm}{\begin{matrix}}
\newcommand{\emm}{\end{matrix}}

\def\mb#1{\mathbf{#1}}

\newcommand{\mA}{\mathcal{A}}
\newcommand{\mB}{\mathcal{B}}
\newcommand{\mZ}{\mathcal{Z}}
\newcommand{\mC}{\mathcal{C}}
\newcommand{\mD}{\mathcal{D}}

\newcommand{\mfC}{\mathfrak{C}}

\DeclareMathOperator{\FPdim}{FPdim}
\DeclareMathOperator{\lcm}{lcm}

\DeclareMathOperator{\Rep}{Rep}
\DeclareMathOperator{\Vect}{Vec}

\DeclareMathOperator{\defects}{defects}

\def\ket#1{\left|#1\right>}

\def\matrixTwo#1#2#3#4{\left(\begin{array}{cc}
    #1 & #2 \\ #3 & #4
\end{array}\right)}

\def\wloop#1{
\scalebox{0.7}{
\begin{tikzpicture}[baseline=-0.65ex,scale=1.3]
 \draw (0,0) ellipse [x radius=10pt, y radius=4pt];
 \draw [->,thin] (0,-4pt)--(2pt,-4pt);
 \node at (4pt,-8pt) {$#1$};
\end{tikzpicture}
}}

\def\wloopVtc#1{
\scalebox{0.9}{
\begin{tikzpicture}[baseline=-0.65ex,scale=1]
 \draw (0,0) ellipse [x radius=4pt, y radius=10pt];
 \draw [->,thin] (4pt,0pt)--(4pt,2pt);
 \node at (9pt,0pt) {$#1$};
\end{tikzpicture}
}}

\def\wline#1{
\scalebox{0.7}{
\begin{tikzpicture}[baseline=-0.65ex,scale=1.3]
 \draw (0,-10pt)--(0,10pt);
 \draw [->,thin] (0,0)--(0,2pt);
 \node at (4pt,-4pt) {$#1$};
\end{tikzpicture}
}}

\def\wlineLabes#1#2#3{
\scalebox{0.9}{
\begin{tikzpicture}[baseline=-0.65ex,scale=1]
 \draw (0,-10pt)--(0,10pt);
 \draw [->,thin] (0,0)--(0,2pt);
 \node at (6pt,-12pt) {$#2$};
 \node at (5pt,0pt) {$#1$};a
 \node at (6pt,10pt) {$#3$};
\end{tikzpicture}
}}

\makeatletter
\newcommand*{\Relbarfill@}{\arrowfill@\Relbar\Relbar\Relbar}
\newcommand*{\xeq}[2][]{\ext@arrow 0055\Relbarfill@{#1}{#2}}
\makeatother

\begin{document}

\title{A Defect Verlinde Formula}

\author{Ce Shen}
\email{scbebetterme@gmail.com }
\author{Ling-Yan Hung}
\email{lyhung@fudan.edu.cn}
\affiliation{State Key Laboratory of Surface Physics, Fudan University, 200433 Shanghai, China}
\affiliation{Department of Physics and Center for Field Theory and Particle Physics, Fudan University, Shanghai 200433, China}
\affiliation{Institute for Nanoelectronic devices and Quantum computing, Fudan University, 200433 Shanghai, China}

\date{January 23, 2019}

\begin{abstract}
We revisit the problem of boundary excitations at a topological boundary or junction defects between topological boundaries in non-chiral bosonic topological orders in 2+1 dimensions. Based on physical considerations, we derive a formula that relates the fusion rules of the boundary excitations, and the ``half-linking'' number between condensed anyons and confined boundary excitations. This formula is a direct analogue of the Verlinde formula. We also demonstrate how these half-linking numbers can be computed in explicit Abelian and non-Abelian examples. As a fundamental property of topological orders and their allowed boundaries, this should also find applications in finding suitable platforms realizing quantum computing devices. 
\end{abstract}
\maketitle


Topological orders contain many intricate structures, such as fusion and braiding among anyons.  This is a topic that has been studied in depth in the past few decades. More recently, it is realized that topological interfaces connecting different topological orders contain various interesting structures \cite{kong_anyon_2014, Lan:2014uaa}. These interfaces can be described by the physics of anyon condensation. Excitations localized at the interface  (confined anyons) satisfy interesting (and generically non-commutative) fusion rules. These objects are also intimately related to topological defect lines in CFT's, which have been extensively studied for example in \cite{petkova_generalised_2001,Petkova:2001zn,Gaberdiel:2002qa,fuchs_fusion_2008,Gaiotto:2014lma,Chang:2018iay}. They form a fusion tensor category that does not have a well-defined braiding structure, but they do admit a non-trivial ``half-linking'' with ``condensed anyons''.

 Eliciting their properties gives us important new insights into the fundamental structure of topological field theories and the relationships between them.

Topological orders have also found applications in quantum computing, given their robustness against decoherence, and it has been proposed in the literature that topological defects might in fact be a more convenient candidate to realize universal computing \cite{cong_topological_2017}.  Therefore, the fusion and ``half-braiding'' properties mentioned above acquire practical significance. 

In this paper, we find that the set of physical data -- namely the fusion rules of the defects and ``half-linking'' -- are related in a way directly analogous to the Verlinde formula. We consider confined anyons in a given boundary condensate, which we subsequently generalized to confined anyons localized at the junction between two different boundary condensates.  This would have applications both in the study of interfaces in CFT, and also experimental realizations of defect based topological quantum computations.


Gapped boundaries can be described by anyon condensation \cite{bais_broken_2002,bais_condensate-induced_2009, 2012CMaPh.313..351K, kong_anyon_2014, Lan:2014uaa,hung_generalized_2015,hung_ground_2015}. For a given non-chiral bulk phase $\mB$ in 2+1 dimensions, there could be multiple different gapped boundaries,  each characterized by a distinct pattern of anyon condensation. 

The bulk phase is describable by a semi-simple modular fusion tensor category $\mB$. The basic physical data is the distinct simple topological sectors, or anyons $a$ with quantum dimension $d_a$. These sectors can fuse, i.e. 
\be
a\otimes b = \oplus_c N_{ab}^c \,c ,
\ee
and these fusion rules are associative. The self-statistics of the anyons are encoded in the eigenvalues of the so called modula $T$ matrix \cite{kirillov_lectures_2001}, and their mutual statistics are encoded in the modular $S$ matrix, $S_{ab}$ \cite{kirillov_lectures_2001}.
The Verlinde formula is a renowned relation between $N_{ab}^c$ and $S_{ab}$ \cite{verlinde_fusion_1988},
\be \label{Verlinde0}
N_{ab}^c = \sum_{d} \frac{S_{ad}S_{bd}(S^{-1})^{ dc}}{S_{0d}}.
\ee 

The most important physical data characterizing a given pattern of anyon condensation corresponding to a gapped boundary is the set of condensed anyons $c \in \mC$. They form a Lagrangian sub-algebra in the bulk phase $\mB$. The condensate would behave like the trivial sector in the condensed phase. In the case of gapped boundary, the condensed phase is the trivial topological order with only one (trivial) sector. Anyons not belonging to the condensate are ``confined'' and they would correspond to boundary excitations as they approach the boundary. Not every confined anyon in the bulk correspond to a distinct boundary excitation, since they are identified if they are related by fusion with one of the condensed anyons $c$. To be precise, the relation between bulk anyons $a$ and boundary excitations $x$ can be expressed using the $W$ matrix \cite{bais_broken_2002,bais_condensate-induced_2009}
\be \label{Wmatrix}
a = \oplus_x W_{a x} x,
\ee
where $W_{ax}$ is a positive integer which gives the ``multiplicity'' of $a$ decomposing into $x$ \footnote{This should be contrasted with the $W$ matrix discussed for example in \cite{Lan:2014uaa}, in which the $x$ indices contain only ``unconfined'' sectors. }. If $a$ is part of the condensate $\mC$, then $W_{a0} \neq 0$. And it follows that
\be
\mC = \oplus_c W_{c0} c.
\ee

The decomposition (\ref{Wmatrix}) commutes with fusion. i.e.
\be \label{decompN}
\sum_c N_{ab}^c W_{cz} =\sum_{x,y} W_{ax}  W_{by}  n_{xy}^z,
\ee
where $n_{xy}^z$ is the fusion coefficient of the boundary excitations corresponding to confined anyons in the condensate.



\begin{figure}[b]
    \includegraphics[width=0.45\textwidth]{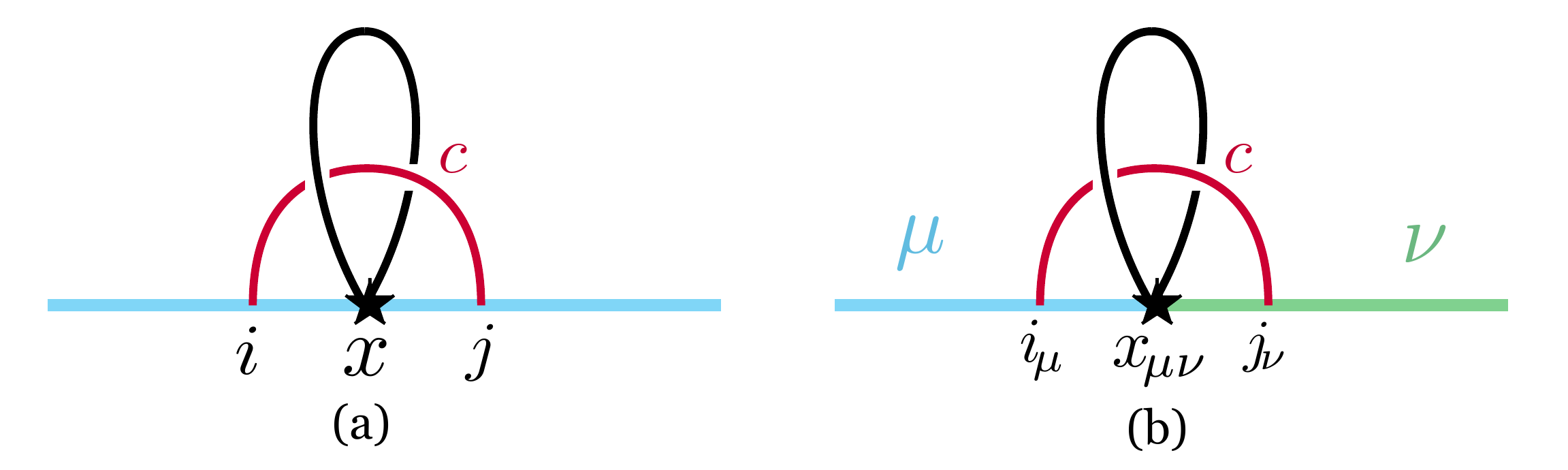}
   
    \caption{(a): Half-linking $\gamma_{x \, c _{({i,j})}}$ with boundary excitation $x$ and condensed anyon $c$. Here $i$ and $j$ label condensation channel. (b): Half-linking $\gamma^{(\mu|\nu)}_{x \, c _{(i_\mu,j_\nu)}}$ with defect $x_{\mu\nu}$ living at the junction between two gapped boundaries $\mu$ and $\nu$, while the shared condensed anyon $c\in\mC_{\mu}\cap\mC_{\nu}$.}
    \label{fig:gamma_matrix}
\end{figure}


Boundary excitations are confined anyons that are stuck at the boundary, not capable of getting past to the condensed phase without creating a trail of excitations. Condensed anyons on the other hand can freely pass through the boundary. This suggests a natural topological number characteristic of the boundary condensate. Namely,  consider pair-creating a boundary anyon $x$. Now a condensed anyon $c$ is created from the boundary, moved around $x$ in the bulk, and finally annihilated at the boundary.
Then we also annihilate the pair of $x$. (see Fig.\ref{fig:gamma_matrix} a) This is, up to normalization, closely related to the {\it half-linking} mentioned in the context of gapped boundaries of  Abelian Chern-Simons theory \cite{kapustin_topological_2011}. Here we are generalizing the discussion to non-Abelian systems, and we denote this quantity by 
\be \label{gamma}
\gamma_{x \, c _{({i,j})}}.
\ee 
Here there are extra indices $i,j$ if $W_{c0} \ge 2$. Then one has to specify the ``condensation channel'' of the condensed anyon $c$, i.e.   $1 \leq i \leq W_{c0}$, where it is created and annihilated at the boundary.  We note that  \cite{hung_ground_2015,Lan:2014uaa,cong_topological_2016}
\be \label{cylinder_deg1}
\sum_{c \in \mB} W_{c0}W_{c0} = N_x, 
\ee
where $N_x$ is the number of distinct simple boundary excitations.  This is a necessary condition for $\gamma_{x \, c _{({i,j})}}$ to be invertible as an $N_x \times N_x$ matrix.
Moreover, given two different gapped boundaries labelled $\mu,\nu$ which are characterized by $W^{(\mu)}$ and $W^{(\nu)}$ respectively, the number $N_{\mu\nu}$ of distinct species of anyons that are localized at the junction of the two boundaries is given by
\be  \label{cylinder_deg2}
N_{\mu\nu}= \sum_{c\in \mB} W^{(\mu)}_{c0}W^{(\nu)}_{c0}.
\ee
This is depicted in Fig.\ref{fig:gamma_matrix} b.

\begin{figure}[t]
    \centering
        \includegraphics[width=0.5 \textwidth]{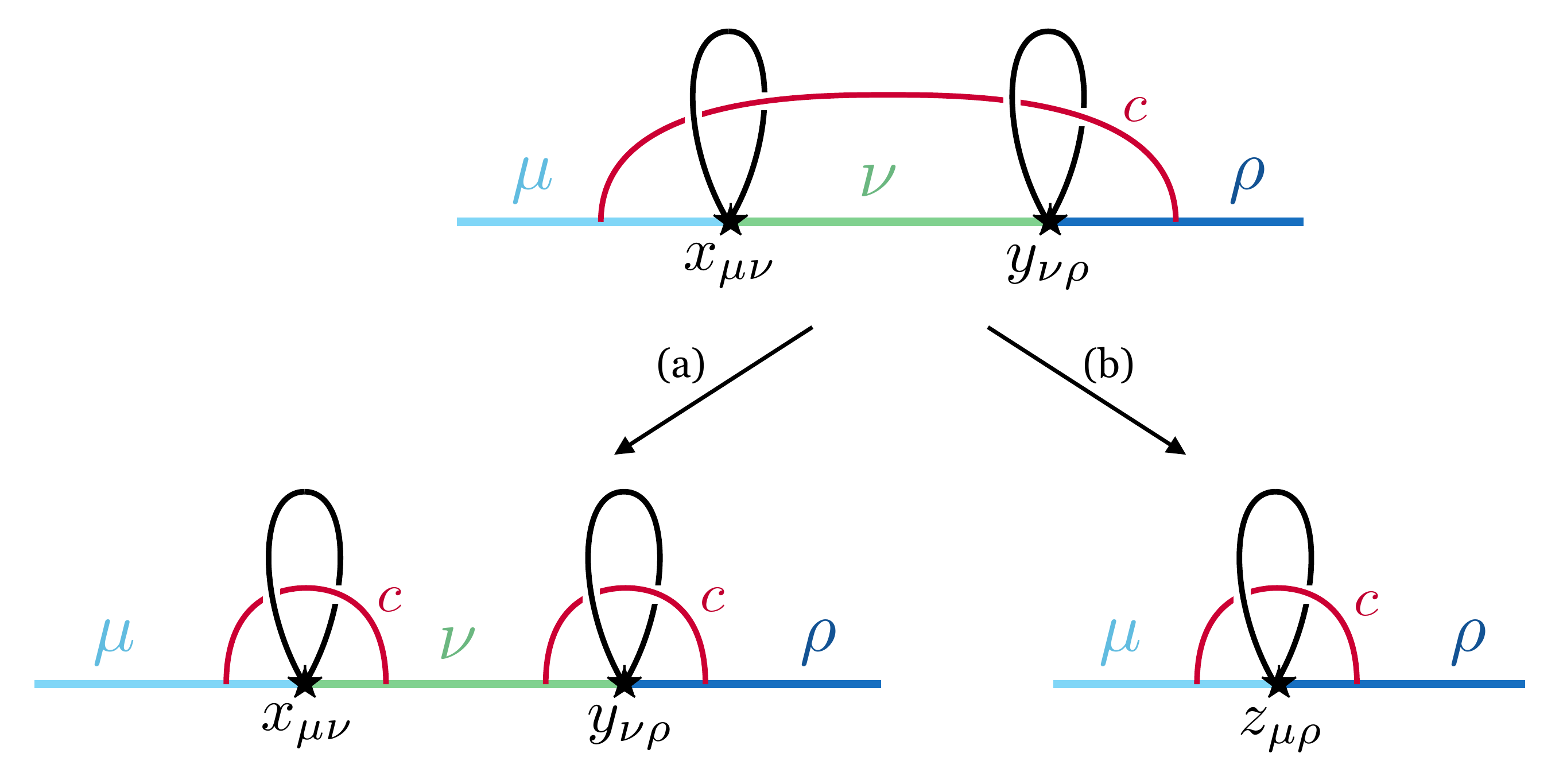}
        \caption{(a): Deforming the condensed $c$ anyon line yields two half-links. (b): Shrinking the middle $\nu$ boundary fuses the two defects $x_{\mu\nu}$ and $y_{\nu\rho}$.}
        \label{fig:fusion}
\end{figure}


By considering the half-linking of a condensed anyon around two boundary excitations $x,y$, before and after the fusion, one can deduce an important relation between the half-linking number and the fusion coefficients $n_{xy}^z$ of boundary excitions. A graphical representation of the processes is given in Fig.\ref{fig:fusion}. Note that we have made use of the fact that the world-line of a condensed anyon can be deformed and breaks up at the boundary.  It leads to the following equation:
\be \label{fuse1}
n_{xy}^z =\sum_c \sum_{i,j,l,k} \gamma_{x \, c _{({i,j})}}(M_{c}^{-1})_{j,l}\gamma_{y \, c _{({l,k})}} \gamma^{-1}_{ c _{(k,i)}\, z}.
\ee 
Here, the half-braiding is an invertible matrix, such that $\gamma^{-1}$ satisfies 
\begin{align}
&\sum_c\sum_{i,j} \gamma_{x \, c _{({i,j})}} \gamma^{-1}_{ \, c _{({j,i})}y} = \delta_{xy}, \\
&\sum_{x}\gamma^{-1}_{ c' _{({j',i'})}\,x}  \gamma_{x \, c _{({i,j})}}= \delta_{c,c^{\prime}} \delta_{i,i'} \delta_{j,{j'}}.
\end{align}

We have introduced a proportionality matrix $(M_{c}^{-1})_{j l}$ which is closely related to the {\it boundary 3$j$-symbols} describing the deformation of condensed anyon lines so that it breaks up at the boundary \cite{cong_topological_2016}. This coefficient can be determined in terms of the half-linking matrices for this boundary. Consider winding a condensed anyon $c$ around vacuum $x=y=0$ .  Then (\ref{fuse1}) implies
\be \label{Msol}
(M_{c})_{\, {i} j} = \gamma_{0c_{(i, j)}},
\ee
where  $ \gamma_{0 \, c _{({i,j})}}$ is taken to be invertible as a matrix with indices ${i}, {j}$.\footnote{From the examples we have worked with, it appears that $M_c$ is proportional to $\delta_{i j}$, which is explicit in the examples built from gauge theories. We also expect that to be a natural result as we break an anyon line into two at the boundary.}

This is the analogue of the Verlinde formula for boundary excitations.  We note that this formula is closely related to the discussion of defect operators in the context of CFT, where boundary excitations correspond to defect line operators in a given modular invariant CFT \cite{petkova_generalised_2001,fuchs_fusion_2008}. This modular invariant is in direct correspondence with the Lagrangian sub-algebra characterizing the gapped boundary. Here, we arrive at the result via a different route that is based on the physical process in the 2+1 dimensional system.  

This result however, can be further generalized. Consider two gapped boundaries $\mu,\nu$ joining at a junction, with corresponding sets of boundary excitations $X_{\mu}$ and $X_\nu$ respectively. Among those confined anyons some are confined in both phases $\mu$ and $\nu$. They are thus localized at the junction connecting the two boundaries $\mu,\nu$. We denote the set of excitations by ${X_{\mu\nu}}$. Among condensed anyons in $\mu,\nu$, there are also shared condensed anyons $c\in \mC_{\mu}\cap \mC_{\nu}$ that can be freely created and annihilated at both boundaries. We can thus define a half-linking between $c$ and $x \in X_{\mu\nu}$, 
\be
\gamma^{({\mu}|{\nu})}_{xc_{(i_\mu,i_\nu)}}, \,\, 1\leq i_{\alpha}  \leq W^{(\alpha)}_{c 0}, \,\,\alpha \in \{\mu,\nu\},
\ee

which reduces to (\ref{gamma}) when $\mu=\nu$.

Now consider three adjacent gapped boundaries $\mu,\nu,\rho$, where $\mu$ is connected to $\nu$ and $\nu$ connected to $\rho$.  
Since these boundaries are gapped, we can adiabatically compress the sizes of these boundaries. Therefore, excitations localized at different junctions $x_{\mu\nu}$ and $y_{\nu\rho}$ can also be fused to obtain localized excitations between $\mu$ and $\rho$, as the $\nu$ boundary is compressed to vanishing size. i.e. the fusion rules are generalized to include
\be
x \otimes y = \oplus_{z}  n_{x \,y}^{z} \, z,\,\, x \in X_{\mu\nu},\,\,y \in X_{\nu\rho},\,\, z\in X_{\mu\rho}
\ee

Then consider the half-link of a condensed anyon $c\in \mC_{\mu}\cap \mC_{\rho}$ around $x$ and $y$. One can either deform the condensed anyon line $c$ so that it becomes two half-links, separately around $x $ and $y$; or we can first fuse $x$ and $y$ before computing the half-linking number with the fusion product. This is illustrated in Fig.\ref{fig:fusion} . This gives another relation between the half-linking numbers and the fusion coefficients: 
\begin{align} \label{fuse2}
&n_{x \,y}^{z}   \nonumber \\
&=\sum_c\sum_{i_{\mu},i_{\nu}, i_{\nu}^{\prime},i_{\rho}} \gamma^{(\mu|\nu)}_{x\,c_{(i_\mu, i_\nu)}} (M^{\nu}_c)^{-1}_{i_{\nu}i_{\nu}^{\prime}}
\gamma^{(\nu|\rho)}_{y\,c_{(i_{\nu}^{\prime},i_{\rho})}} (\gamma^{(\mu|\rho)})^{-1}_{c_{({i_{\rho},i_{\mu}})}z} \nonumber \\
&=  \sum_c \gamma^{(\mu|\nu)}_{xc}\, (M^{\nu}_c)^{-1} \,\gamma^{(\nu|\rho)}_{yc}\, (\gamma^{(\mu|\rho)})^{-1}_{c z},
\end{align}
where we have included a subscript $\nu$ for the matrix $M_c$, and simplified our notation in the last line to keep the condensation channel indices $i_{\alpha}$ implicit.
Equation (\ref{fuse1},\ref{fuse2}) are the main results of this paper. Equation (\ref{fuse2}), together with accompanying examples that we will discuss, are to the best of our knowledge, also appearing for the first time in the mathematics literature. 


The formula above is only useful if we can compute these matrices. We make an observation here. Consider putting a topological order $\mB$ on a cylinder, with the top boundary labeled $\mu$, and the bottom boundary labeled $\nu$. The ground state degeneracy is given by $N_{\mu\nu}$ in (\ref{cylinder_deg2}). There are two possible sets of basis states, namely states corresponding to confined anyons winding the non-contractible loop, 
labeled by $\ket{\wloop{x}}$, and condensed anyons connecting the top and bottom boundaries labeled by $\ket{\wlineLabes{c}{i_{\nu}}{i_{\mu}} }$. One finds that the change of basis is effected by the half-linking numbers:
\be \label{transbas}
\ket{\wloop{x}} = \sum_{c\in\mC_{\mu}\cap\mC_{\nu}}\sum_{i_{\mu},i_{\nu}}\gamma^{(\mu|\nu)}_{x\,c_{(i_{\mu},i_{\nu})}} \ket{\wlineLabes{c}{i_{\nu}}{i_{\mu}} }.
\ee
This gives us a very practical way of computing these matrices in explicit models, such as Abelian Chern-Simons theories, and lattice models. We discover useful relations from these exercises that can be applied in other more general situations.  Moreover, given that it is a basis transformation, it is expected to be a unitary transformation. This is indeed the case in all the examples we encountered. We will therefore in the following write
\be
(\gamma^{(\mu|\nu)})^{-1} = \overline{\gamma^{(\mu|\nu)}}.
\ee


 We will first consider the case of a single boundary where the corresponding anyon condensation channel multiplicity $W_{c0}\leq 1$ for a given topological boundary. In this case, we can provide an alternative proof of (\ref{fuse1}) where no extra multiplicity indices $c_i$ are needed.
Using (\ref{decompN}, \ref{Verlinde0}), we obtain
\be \label{fuseW1}
n_{xy}^z = \sum_{c} \frac{V_{xc} V_{yc} V^{-1}_{cz}}{S_{0c}}, \,\, V^{-1}_{cx} = \sum_{a} \overline{S_{{a}c}}W_{ax},
\ee
producing an alternate form of the defect Verlinde formula for a given boundary.
Suppose
\be \label{relategV}
\frac{\gamma_{xc}}{\gamma_{1c}} = V^{-1}_{cx}, \qquad \gamma_{1c} = \sqrt{S_{1c}}.
\ee
Substituting into (\ref{fuseW1}), we recover (\ref{fuse1}).
The relation (\ref{relategV}) is an empirical observation based on Abelian Chern-Simons theories, the quantum double model $D(S_3)$, and also generic tensor product theories $\mB = \mathcal{D} \boxtimes \bar{\mathcal{D}}$ for $\mathcal{D}$ some modular tensor category, and $\bar{\mathcal{D}}$ its time-reversal. In this last case, ``diagonal-condensation'' always gives an allowed gapped boundary, where every anyon of the form $(a\bar{a})$ condenses, for $a\in \mathcal{D}$. The half-linking matrix which defines the transformation (\ref{transbas}) would coincide with the modular matrix $S^{\mathcal{D}}$ of $\mathcal{D}$. This follows from the fact that the cylinder with diagonal condensates on both edges can be unfolded into the phase $\mathcal{D}$ covering a torus.
Given these infinite classes of theories, we conjecture that (\ref{relategV}) is true in general where $W_{c0}\leq 1$.

Explicit illustration by examples are relegated to the appendix.  

More interestingly, in non-Abelian systems, one could have boundaries characterized by anyon condensation with $W_{c0}> 1$. We illustrate this using the quantum double model $D(G)$ \cite{kitaev_fault-tolerant_2003}, which is a gauge theory with gauge group $G$.  Different anyonic sectors correspond to different conjugacy classes and irreducible representations of their centralizers. We will particularly consider non-Abelian groups. Among the gapped boundaries, one corresponds to the so-called {\it electric condensate}, where anyons corresponding to all the representations of $G$ condense. A representation $R$ of dimension $d_R$ would carry a condensation multiplicity $W_{R0} = d_R$ \cite{Lan:2018vjb}. There are $|G|$ types of defects (confined anyons) in the condensate, corresponding to the magnetic charges of the gauge group $G$, which are labelled simply by the group elements $g\in G$. While magnetic anyons in the bulk phase are labelled by conjugacy classes of the gauge group, they split up into individual group elements in an electric condensate. On a cylinder with electric boundaries, the two sets of basis states discussed in (\ref{transbas}) are 
\begin{align}
\ket{\wloop{x}} &\rightarrow |g\rangle,\qquad \qquad g\in G \\ 
\ket{\wlineLabes{c}{{i}}{{j}}}
&\rightarrow  |R, R_i,R_j\rangle, \,\, 1\leq R_{i,j}\leq d_R.
\end{align}  
In this case, one can readily work out the transformation matrix between these basis states, which, as expected, is the direct analogue of the transformation between group and representation basis states in a lattice gauge theory \cite{2009PhRvB..80o5136B}:
\be \label{group_electric}
|g\rangle = \sum_{R,R_i,R_j} \sqrt{\frac{d_R}{|G|}} \rho^R_{R_j, R_i }(g^{-1}) |R, R_i, R_j\rangle. 
\ee

Now, substituting into (\ref{fuse1}), we have
\begin{align}
N_{g_1 g_2}^{g_3} &= \sum_{R} \frac{d_R}{|G|}  \chi^R(g_1^{-1}g^{-1}_2g_3)  \\
&= \delta_{g_2 \times g_1, g_3}.
\end{align}
We have used (\ref{Msol}) above, which reduces here to
\be
 M_R = \sqrt{\frac{d_R}{|G|}} \mathbb{I}_R.
\ee
The fusion of the defects coincide with the group product rule, which gives a natural realization of a non-commutative fusion ring envisaged in the CFT literature \cite{Petkova:2001zn}. Invertible Verlinde lines in RCFT are expected also to satisfy a fusion rule that is equivalent to group multiplication \cite{Chang:2018iay}. 
We derive these rules using the defect Verlinde formula here. 

We would now like to discuss explicit examples realizing (\ref{fuse2}). 

To be explicit, there is a class of Abelian theories whose boundaries can be easily described.  These Abelian theories are equivalent to the quantum doubles $D(G)$ where $G$ is a finite Abelian group. The quantum dimension of the theory is given by $D = |G|$.
Each boundary is characterized by an abelian subgroup $K\subset G$, which is in direct correspondence with a set of condensed anyons.
With details left in the appendix, we find that the half-linking matrix at a junction between two boundaries characterized by two subgroups $K_{\mu,\nu}$ is given by
\begin{align}
    \label{eq:gamma_matrix}
   \gamma_{xc}^{(\mu|\nu)}
    =\frac{1}{\sqrt{D}}\frac { \sqrt { \left| K _ { \mu } \right| \cdot\left| K _ { \nu } \right| } } { \left| K _ { \mu } \cap K _ { \nu } \right| }\tilde{S}_{x^{\mB}\,c}.
\end{align}
The matrix $\tilde{S}$ is proportional to the bulk S-matrix, 
\be
\tilde{S}_{x\, c} = \exp(2\pi i \mb{l}_c^{T}K^{-1}(\mb{l}_x)),
\ee
and $\mb{l}_{c}, \mb{l}_x$ are (2-component) integer charge vectors corresponding to the topological sector $c$ and $x$ respectively. In (\ref{eq:gamma_matrix}),  $x^{\mB}$ is a representative bulk anyon chosen satisfying $W^{\mu}_{x^{\mB} x} = W^{\nu}_{x^{\mB} x}=1,\,\, x\in X_{\mu\nu}$. Substituting into (\ref{fuse2}) it recovers the fusion rule
\begin{align}
    \label{eq:abelian_fusion_rule}
 & \delta_{\mb{l}_{x}+\mb{l}_{y},\mb{l}_{z}} \equiv n_{x y}^{z}=
 \sum_{c\in \mC_\mu\cap \mC_\nu \cap \mC_\rho}{{\gamma}_{xc}^{(\mu|\nu)}  {M^{(\nu)}_c}^{-1} {\gamma}_{yc}^{(\nu|\rho)}\overline{{\gamma}_{zc}^{(\mu|\rho)}}}, \nonumber \\
 & x\in X_{\mu\nu},\,\, y \in X_{\nu\rho},\,\, z \in X_{\mu\rho}
\end{align}
where $C_{\mu}$ is the set of condensed anyons of the boundary $\mu$.

Let us give a non-Abelian example based on the quantum double $D(S_3)$.
We will illustrate the fusion of defects at junctions between two gapped boundaries characterized by the condensate $\mA_3=A\oplus C\oplus D$ and $\mA_4=A\oplus F\oplus D$. (Another interesting junction between $\mA_1=A\oplus B\oplus 2C$ and $\mA_3=A\oplus C\oplus D$ is relegated to the appendix.) 
The half-linking matrices between various $S_3$ boundaries are listed in the appendix. 

For $\mA_3|\mA_4|\mA_3$ defects fusion, 
we calculate the following fusion coefficients from (\ref{eq:S3_ACD_AFD}, \ref{eq:S3_ACD_ACD}, \ref{eq:S3_AFD_AFD}) 
\begin{eqnarray}
    \label{eq:A3A4A3_fusion}
    n_{xy}^z=\sum_{c\in \{A,D\} }\frac{\gamma_{xc}^{(3|4)} \gamma_{yc}^{(4|3)} \overline{\gamma_{zc}^{(3|3)}} } { \gamma_{ 0 c}^{(4|4)} }.
\end{eqnarray}
Here the $M_c$ matrix is simply a number $\gamma_{ 0 c}^{(4|4)}$.
We can derive the fusion rules:
\begin{eqnarray}
    \{A\}^{(3|4)} \otimes  \{A\}^{(4|3)} = \{A\}^{(3|3)} \oplus \{F\}^{(3|3)}\nonumber\\
    \{A\}^{(3|4)} \otimes  \{B\}^{(4|3)} = \{B\}^{(3|3)} \oplus \{F\}^{(3|3)}\nonumber\\
    \{B\}^{(3|4)} \otimes  \{B\}^{(4|3)} = \{A\}^{(3|3)} \oplus \{F\}^{(3|3)}
\end{eqnarray}
The fusion involving the defect $\{A\}$ was obtained in \cite{cong_defects_2017}, based on a heuristic method.
The complete fusion rules are now derived using the defect Verlinde formula.

\section{Conclusion}
In this paper, we have studied the boundary excitations and junction excitations at and between topological boundaries. 
We find that the fusion algebra of these boundary excitations can be naturally connected to the ``half-linking'' matrix, in direct analogy 
of the Verlinde formula. We derive an explicit expression of this defect Verlinde formula, and presented examples, including both Abelian and non-Abelian ones.
This should find numerous applications in recovering fusion algebra of defects in topological orders, which are hopeful candidates for construction of viable quantum computing devices \cite{kitaev_fault-tolerant_2003,Barkeshli:2012pr,cong_topological_2017}.

\section{Acknowledgements}
We thank Laurent Freidel, Davide Gaiotto, Matthias Gaberdiel, Liang Kong, Yidun Wan, Xiao-Gang Wen and Gabriel Wong  for discussions. Part of this work is done
during the authors' visit to Perimeter Institute as part of the Emmy-Noether Fellowship programme. 
 LYH acknowledges the support of Fudan University and the Thousands Young Talents Program. This work is supported by the NSFC grant number 11875111.

\appendix

\section{Appendix}

\section{A brief review of Abelian Chern-Simons theories and their gapped boundaries}

The Abelian Chern-Simons theory is described with action given by
\be
S = \frac{1}{4\pi}\int K_{IJ} A^I dA^J,
\ee 
where $K_{IJ}$ is a symmetric quadratic form with total signature $0$ as a necessary condition to support gapped boundaries.
It is equivalent to the quantum doubles $D(G)$ for some finite Abelian group $G$.  
For example the $K$ matrix 
\begin{align}
    \label{eq:abelian_K_matrix}
   K= \matrixTwo{0}{N}{N}{0}
\end{align}
describes the topological order $D(\Z_N)$, which will be the class of Abelian examples we consider here. 

In Abelian theory each anyon has quantum dimension $1$, and the total quantum dimension of the system is $|G|$.
Each gapped boundary is characterized by an Abelian subgroup of $G$. Given this subgroup, the boundary condensate can be worked out via a generic procedure in \cite{beigi_quantum_2011} by calculating the characters.

From the general structure of finite Abelian group, we can assume that $G$ is a cyclic group without loss of generality. Then the subgroups of $G$ are also cyclic groups. Suppose $N$ is a positive integer and $m,n$ are its positive divisors generating subgroup $K_{\mu}$ and $K_{\nu}$ respectively, we take
\begin{align}
    G&=\{0,1,2,\dots,N-1\}\cong \Z_N, \nonumber\\
    K_{\mu}&=\{0,m,2m,\dots,N-m\}\cong \Z_{N/m}, \nonumber\\
    K_{\nu}&=\{0,n,2n,\dots,N-n\}\cong \Z_{N/n}. 
\end{align}
Then $K _ {\mu} \cap K _ { \nu }$ is generated by $\lcm(m,n)$, and $G/(K_{\mu}K_{\nu})$ is generated by ${N}/{\gcd(m,n)}$.

Each topological sector is represented by a (here, two-component) integer valued charge vector $\mb{l}$. The representation is not unique, and are identified under
\be
\mb{l} \sim \mb{l} + K. \mb{m}, \,\, \mb{m} \in \mathbb{Z}^2.
\ee
Each condensate here is generated by two $2-$component vectors.
\begin{align}
\mC_{\mu}&=\left\langle \left( \begin{array} { l } { m } \\ { 0 } \end{array} \right) ,~ \left( \begin{array} { c } { 0 } \\ { N / m } \end{array} \right) \right\rangle, \nonumber\\
\mC_{\nu}&=\left\langle \left( \begin{array} { l } { n } \\ { 0 } \end{array} \right) ,~ \left( \begin{array} { c } { 0 } \\ { N / n } \end{array} \right) \right\rangle.
\end{align}
From which is found the subgroup of shared condensed anyons
\begin{align} 
    \label{eq:shared_condensed}
\mC_{\mu}\cap\mC_{\nu}\cong G/(K_{\mu}K_{\nu})\times(K_{\mu}\cap K_{\nu}).
\end{align}
A finite Abelian group can be decomposed into direct products of cyclic groups, so (\ref{eq:shared_condensed}) is correct for an arbitrary Abelian group $G$, although it's derived from a cyclic group $G\cong\Z_N$.

For an Abelian theory, all defects between boundary $\mu$ and $\nu$ have identical quantum dimension \cite{Barkeshli:2012pr, barkeshli_classification_2013,cong_topological_2016}
\begin{align}
    \label{eq:defect_dimension}
    \FPdim(K_{\mu}|K_{\nu})=\frac { \sqrt { \left| K _ { \mu } \right| \cdot\left| K _ { \nu } \right| } } { \left| K _ { \mu } \cap K _ { \nu } \right| },
\end{align}
while the number of defects is 
\begin{align}
    \label{eq:number_defects}
\#(\defects)=\left| K _ { \mu } \backslash G / K _ { \nu } \right| \cdot \left| K _ { \mu } \cap K _ { \nu } \right| .
\end{align}

In Abelian case the double coset $K _ { \mu } \backslash G / K _ { \nu}$ is isomorphic to the quotient group $G/(K_{\mu}K_{\nu})$.
Comparing (\ref{eq:shared_condensed}) with (\ref{eq:number_defects}), we've shown explicitly that, in Abelian Chern-Simons theory, the number of boundary defects is equal to the number of shared condensed anyons.

As pointed out in the main text, the half-linking matrix is equivalent to the transformation matrix of two sets of ground state basis on a cylinder. The two sets of basis are constructed as follows.The top $\mu$ (resp. bottom $\nu$) boundary allows the condensed anyons $\mC_{\mu}$ (resp. $\mC_{\nu}$) to escape from the bulk. Only anyons confined wrt to both boundaries can wind around the non-contractible loop of the cylinder, forming a basis of the ground state subspace. 
On the other hand, the non-trivial ground state degeneracy can also be labeled by the shared condensed anyon connecting the top and bottom boundaries. 
The basis transformation between these two sets of basis is discussed in (\ref{transbas}).
Note that $W_{0c}\leq 1$ in Abelian cases.
In which case, (\ref{transbas}) reduces to
\begin{align}
    \ket{\wloop{x}}=\sum_{c\in\mC_{\mu}\cap\mC_{\nu}}\gamma^{(\mu|\nu)}_{xc}\ket{\wline{c}}.
\end{align}
The confined anyon is morally the defect $x$ localized at the junction between boundary $\mu$ and $\nu$, while the shared condensed anyon line is the $c-$semicircle as in Fig.\ref{fig:gamma_matrix}. Therefore we can obtain the half-linking matrix by calculating the basis transformation on a cylinder.

The trivial loop state is a democratic average of the line basis. Including a normalization factor, we have
\begin{align}
    \ket{\wloop{0}}&=\frac{1}{\sqrt{\left| K _ { 1 } \backslash G / K _ { 2 } \right| \cdot \left| K _ { 1 } \cap K _ { 2 } \right|}}\sum_{c}\ket{\wline{c}} \nonumber\\
    &=\frac{\FPdim(K_{\mu}|K_{\nu})}{\sqrt{D}}\sum_c\ket{\wline{c}}.
\end{align}
    
A general loop state is obtained by acting the Wilson loop operator on the trivial loop state:
\begin{align}
    \ket{\wloop{x}}&=\hat{\wloop{x}}\ket{\wloop{0}}\nonumber\\
    &=\frac{\FPdim(K_{\mu}|K_{\nu})}{\sqrt{D}}\sum_c\hat{\wloop{x}}\ket{\wline{c}}\nonumber\\
    &=\frac{\FPdim(K_{\mu}|K_{\nu})}{\sqrt{D}}\sum_c\frac{S_{xc}}{S_{0c}}\ket{\wline{c}}\nonumber\\
\end{align}
Where in the last line use has been made of the fact that $\ket{\wline{c}}$ is an eigenstate of $\hat{\wloop{x}}$ with eigenvalue $\frac{S_{xc}}{S_{0c}}$ \cite{cong_topological_2016}. 
The basis transformation matrix, and hence the the half-linking matrix, for the Abelian Chern-Simons theory is given by
\begin{align}
    \label{eq:abelian_gamma_matrix}
    \gamma_{xc}^{(\mu|\nu)}
    &=\frac{1}{\sqrt{D}}\FPdim(K_{\mu}|K_{\nu})\cdot\tilde{S}_{xc},
\end{align}
where $\tilde{S}_{xc} = \exp(2\pi i \mb{l}_c^{T}K^{-1}\mb{l}_x)$ is proportional to the bulk $S$-matrix, and the quantum dimension is given in (\ref{eq:defect_dimension}). 
It's easy to check that this half-linking matrix is unitary.

Consider three adjacent boundaries $\mu,\nu,\rho$, and an anyon line $c$ that is shared by all three condensates.
As a direct generalization of (\ref{eq:shared_condensed}),
\begin{align}
\mC_{\mu}\cap \mC_{\nu} \cap \mC_{\rho}\cong G/(K_{\mu}K_{\nu}K_{\rho})\times (K_{\mu}\cap K_{\nu}\cap K_{\rho})
\end{align}

Using (\ref{eq:abelian_gamma_matrix}), a direct calculation gives
\begin{align}
    &\sum_{c\in \mC_{\mu}\cap \mC_{\nu} \cap \mC_{\rho} }\frac{\gamma_{xc}^{(\mu|\nu)} \gamma_{yc}^{(\nu|\rho)} \overline{\gamma_{zc}^{(\mu|\rho)}}} {\gamma_{0c}^{(\nu|\nu)}}\nonumber\\
    &=\frac{1}{D}\frac{\FPdim(K_{\mu}|K_{\nu}) \FPdim(K_{\nu}|K_{\rho}) \FPdim(K_{\mu}|K_{\rho})}{\FPdim(K_{\nu}|K_{\nu})}\nonumber\\
    &\qquad\times\sum_{c\in \mC_{\mu}\cap \mC_{\nu} \cap \mC_{\rho} } \tilde{S}_{xc}\tilde{S}_{yc}\overline{\tilde{S}_{xc}} \nonumber\\
    &=\frac{1}{|G|}\frac { \sqrt { \left| K _ { \mu } \right| \cdot\left| K _ { \nu } \right| } } { \left| K _ { \mu } \cap K _ { \nu } \right| } \frac { \sqrt { \left| K _ { \nu } \right| \cdot\left| K _ { \rho } \right| } } { \left| K _ { \nu } \cap K _ { \rho } \right| }\frac { \sqrt { \left| K _ { \mu } \right| \cdot\left| K _ { \rho } \right| } } { \left| K _ { \mu } \cap K _ { \rho } \right| }\nonumber\\
    &\qquad\times |\mC_{\mu}\cap \mC_{\nu} \cap \mC_{\rho}| \cdot \delta_{\mb{l}_x+\mb{l}_y,\mb{l}_z}\nonumber\\
    &=\delta_{\mb{l}_x+\mb{l}_y,\mb{l}_z}\nonumber\equiv n_{xy}^z
\end{align}

This is the defect Verlinde formula for Abelian Chern-Simons theory. Note that $M_c^{\nu}=\gamma_{0c}^{(\nu|\nu)}$ in (\ref{eq:abelian_fusion_rule}).

\section{A brief review of the quantum double $D(S_3)$ and its gapped boundary}
We would like to review here some basic data of the $D(S_3)$ model.
Let $S_3=\langle s=(23),~r=(123) \rangle$.
    The anyons are labeled by $(C,\rho_{\alpha_C})$, where $C$ is a conjugacy class of the group $G=S_3$, and $\alpha_C$ an irrep of the centralizer of $C$. A summary of all the anyons are listed below. (The trivial sector is conventionally denoted ``A''.)
    
    \scalebox{0.88}{
        \begin{tabular}{c|ccc|cc|ccc}
            \hline\hline
             & $A$ & $B$ & $C$ & $D$ & $E$ & $F$ & $G$ & $H$ \\
            \hline
            conjugacy class $W$ & \multicolumn{3}{|c|}{$\{e\}$} & \multicolumn{2}{c|}{$\{s,rs,r^2s\}$} & \multicolumn{3}{c}{$\{r,r^2\}$} \\
            \hline
            centralizer $\cong$ & \multicolumn{3}{|c|}{$S_3$} & \multicolumn{2}{c|}{$Z_2$} & \multicolumn{3}{c}{$Z_3$} \\
            \hline
            irrep $\rho$ of centralizer  & \boldmath$1$ & sign & \boldmath$\pi$ & \boldmath$1$ & \boldmath$-1$ & \boldmath$1$ & \boldmath$\omega$ & \boldmath$\omega^*$ \\
            \hline
            dim($\rho$) & 1 & 1 & 2 & 1 & 1 & 1 & 1 & 1 \\
            \hline
            \makecell{quantum dimension \\ $d=|W|\times$ dim$(\rho)$} & 1 & 1 & 2 & 3 & 3 & 2 & 2 & 2\\
            \hline
            twist $\theta$ & 1 & 1 & 1 & 1 & -1 & 1 & $e^{2\pi i/3}$ & $e^{-2\pi i/3}$\\
            \hline\hline
        \end{tabular} 
    }
    Their fusion rules are given in Table \ref{tab:s3_fusion}
    \begin{table}
        \scalebox{0.7}{
        \begin{tabular}{|c|c|c|c|c|c|c|c|c|}
            \hline
            $\otimes$ & $A$ & $B$ & $C$ & $D$ & $E$ & $F$ & $G$ & $H$ \\ \hline
            $A$ & $A$ & $B$ & $C$ & $D$ & $E$ & $F$ & $G$ & $H$ \\ \hline
            $B$ & $B$ & $A$ & $C$ & $E$ & $D$ & $F$ & $G$ & $H$ \\ \hline
            $C$ & $C$ & $C$ & $A\oplus B\oplus C$ & $D\oplus E$ & $D\oplus E$ & $G\oplus H$ & $F\oplus H$ & $F\oplus G$ \\ \hline
            $D$ & $D$ & $E$ & $D\oplus E$ & \makecell{$A\oplus C\oplus F$ \\ $\oplus G\oplus H$} & \makecell{$B\oplus C\oplus F$ \\ $\oplus G\oplus H$} & $D\oplus E$ & $D\oplus E$ & $D\oplus E$ \\ \hline
            $E$ & $E$ & $D$ & $D\oplus E$ & \makecell{$B\oplus C$ \\$\oplus F\oplus G\oplus H$} & \makecell{$A\oplus C$ \\ $\oplus F\oplus G\oplus H$} & $D\oplus E$ & $D\oplus E$ & $D\oplus E$ \\ \hline
            $F$ & $F$ & $F$ & $G\oplus H$ & $D\oplus E$ & $D\oplus E$ & $A\oplus B\oplus F$ & $C\oplus H$ & $C\oplus G$ \\ \hline
            $G$ & $G$ & $G$ & $F\oplus H$ & $D\oplus E$ & $D\oplus E$ & $C\oplus H$ & $A\oplus B\oplus G$ & $C\oplus F$ \\\hline
            $H$ & $H$ & $H$ & $F\oplus G$ & $D\oplus E$ & $D\oplus E$ & $C\oplus G$ & $C\oplus F$ & $A\oplus B\oplus H$ \\
            \hline
        \end{tabular}
        }
        \caption{Fusion table of $D(S_3)$}
        \label{tab:s3_fusion}
    \end{table}

    The $S$-matrix is given by
    \begin{eqnarray}
        \label{eq:S3_S_matrix}
        S=\frac{1}{6}\left(
        \begin{array}{cccccccc}
           1 & 1 & 2 & 3 & 3 & 2 & 2 & 2 \\
           1 & 1 & 2 & -3 & -3 & 2 & 2 & 2 \\
           2 & 2 & 4 & 0 & 0 & -2 & -2 & -2 \\
           3 & -3 & 0 & 3 & -3 & 0 & 0 & 0 \\
           3 & -3 & 0 & -3 & 3 & 0 & 0 & 0 \\
           2 & 2 & -2 & 0 & 0 & 4 & -2 & -2 \\
           2 & 2 & -2 & 0 & 0 & -2 & -2 & 4 \\
           2 & 2 & -2 & 0 & 0 & -2 & 4 & -2 \\
        \end{array}    
        \right).
    \end{eqnarray}

    \begin{table}
        \centering
        \scalebox{0.78}{
        \begin{tabular}{|C|C|C|C|C|} 
            \hline
            \mfC & \mA_1 & \mA_2 & \mA_3 & \mA_4 \\ \hline
            \makecell{\mA_1=A\oplus B\oplus 2C\\ K_1=\{1\}} & \Vect_{S_3} & \{\sqrt{3},\sqrt{3}\} & \{\sqrt{2},\sqrt{2},\sqrt{2}\} & \{\sqrt{6}\} \\ \hline
            \makecell{\mA_2=A\oplus B\oplus 2F\\ K_2=\Z_3} & \{\sqrt{3},\sqrt{3} \} & \Vect_{S_3} & \{\sqrt{6}\} & \{\sqrt{2},\sqrt{2},\sqrt{2}\} \\ \hline
            \makecell{\mA_3=A\oplus C\oplus D\\ K_3=\Z_2} & \{\sqrt{2},\sqrt{2},\sqrt{2}\} & \{\sqrt{6}\} & \Rep(S_3) & \{\sqrt{3},\sqrt{3} \} \\ \hline
            \makecell{\mA_4=A\oplus F\oplus D\\ K_4=S_3} & \{\sqrt{6}\} & \{\sqrt{2},\sqrt{2},\sqrt{2}\} & \{\sqrt{3},\sqrt{3} \} & \Rep(S_3) \\ \hline
        \end{tabular}
        }
        \caption{Summary of the distinct boundaries labeled by four different condensates $\mA_{1,2,3,4}$, and the quantum dimension of defects/excitations localized between them. This is reproduced from \cite{cong_defects_2017}. The diagonal cells give the fusion category describing the boundary excitations of each type of boundary. }
        \label{tab:s3_multifusion_cat}
    \end{table}

\setlength{\abovedisplayskip}{3pt}
\setlength{\belowdisplayskip}{3pt} 

In the most general case, these half-linking matrices can be obtained by acting the Wilson line and Wilson loop operators on the basis states, which will involve calculating the {\it boundary 3$j$-symbols} defined in \cite{cong_topological_2016}. 

We list here all inequivalent half-linking matrices between $D(S_3)$ boundaries, keeping in mind that the other half-linking matrices can be obtained by $\gamma^{(i|j)}=\gamma^{(j|i)}$ due to the time-reversal symmetry, or by swapping $C$ and $F$ label due to $C\leftrightarrow F$ duality in $D(S_3)$. The defect label is inside a bracket $\{\cdot\}$ to be distinguished from bulk anyons.  \\

$\mA_3=A\oplus C\oplus D/\mA_2=A\oplus B\oplus 2F$ 
\begin{eqnarray}
    \label{eq:S3_ACD_AB2F}
    \gamma^{(3|2)}=1=\frac{1}{\sqrt{6}}~~
    \begin{blockarray}{cc}
        A \\
        \begin{block}{(c)c}
            \sqrt{6} & ~\{A\} \\
        \end{block}
        \begin{block}{cc}
            & \\
        \end{block}
    \end{blockarray}
\end{eqnarray}


The GSD is $1$, meaning there's only one basis state in the ground state subspace, so the basis transformation matrix is trivial.

$\mA_3=A\oplus C\oplus D/\mA_4=A\oplus F\oplus D$
\begin{eqnarray}
    \label{eq:S3_ACD_AFD}
    \gamma^{(3|4)}=\frac{1}{\sqrt{6}}~~
    \begin{blockarray}{ccc}
        A & D \\
        \begin{block}{(cc)c}
          \sqrt{3} & \sqrt{3} & ~\{A\}  \\
          \sqrt{3} & -\sqrt{3} & ~\{B\} \\
        \end{block}
        \begin{block}{ccc}
            & & \\
        \end{block}
        \end{blockarray}
\end{eqnarray}


$\mA_3=A\oplus C\oplus D/\mA_3=A\oplus C\oplus D$
\begin{eqnarray}
    \label{eq:S3_ACD_ACD}
    \gamma^{(3|3)}=\frac{1}{\sqrt{6}}~~
    \begin{blockarray}{cccc}
        A & C & D \\
        \begin{block}{(ccc)c}
            1 & \sqrt{2} & \sqrt{3} & ~\{A\} \\
            1 & \sqrt{2} & -\sqrt{3} & ~\{B\} \\
            2 & -\sqrt{2} & 0 & ~\{F\} \\
        \end{block}
        \begin{block}{cccc}
            & & & \\
        \end{block}
    \end{blockarray}
\end{eqnarray}

$\mA_4=A\oplus F\oplus D/\mA_4=A\oplus F\oplus D$
\begin{eqnarray}
    \label{eq:S3_AFD_AFD}
    \gamma^{(4|4)}=\frac{1}{\sqrt{6}}~~
    \begin{blockarray}{cccc}
        A & F & D \\
        \begin{block}{(ccc)c}
            1 & \sqrt{2} & \sqrt{3} & ~\{A\} \\
            1 & \sqrt{2} & -\sqrt{3} & ~\{B\} \\
            2 & -\sqrt{2} & 0 & ~\{C\} \\
        \end{block}
        \begin{block}{cccc}
            & & & \\
        \end{block}
    \end{blockarray}
\end{eqnarray}
This half-linking matrix is equivalent to the above $\gamma^{(3|3)}$ by swapping $C\leftrightarrow F$, it's presented here so that the $\mA_3|\mA_4|\mA_3$ example(\ref{eq:A3A4A3_fusion}) is better understood.


$\mA_3=A\oplus C\oplus D/\mA_1=A\oplus B\oplus 2C$
\begin{eqnarray}
    \label{eq:S3_ACD_AB2C}
    \gamma^{(3|1)}=\frac{1}{\sqrt{6}}~~
    \begin{blockarray}{cccc}
        A & C^{1} & C^{2} \\
        \begin{block}{(ccc)l}
            \sqrt{2} & 1 & \sqrt{3} & ~~\{A\} \\
            \sqrt{2} & 1 & -\sqrt{3} & ~~\{F,p_1\} \\
            \sqrt{2} & -2 & 0 & ~~\{F,p_2\} \\
        \end{block}
        \begin{block}{cccc}
            & & & \\
        \end{block}
    \end{blockarray}
\end{eqnarray}
There're two condensation channels in the bottom $\mA_1$ boundary, and we have to perform a so called ``idempotent completion'' to label the defects. \cite{cong_topological_2016} \\


$\mA_1=A\oplus B\oplus 2C/\mA_1=A\oplus B\oplus 2C$
\begin{eqnarray}
    \label{eq:S3_AB2C_AB2C}
    \gamma^{(1|1)}=\frac{1}{\sqrt{6}}~~
    \begin{blockarray}{ccccccc}
        A & B & C^{1,1} & C^{1,2} & C^{2,1} & C^{2,2} \\
        \begin{block}{(cccccc)l}
            1 & 1 & \sqrt{2} & 0 & 0 & \sqrt{2} & ~\{e\} \\
            1 & -1 & \sqrt{2} & 0 & 0 & -\sqrt{2} & ~\{s\} \\
            1 & 1 & -\frac{1}{\sqrt{2}} & -\sqrt{\frac{3}{2}} & \sqrt{\frac{3}{2}} & -\frac{1}{\sqrt{2}} & ~\{r\} \\
            1 & -1 & -\frac{1}{\sqrt{2}} & -\sqrt{\frac{3}{2}} & -\sqrt{\frac{3}{2}} & \frac{1}{\sqrt{2}} & ~\{sr\}\\
            1 & 1 & -\frac{1}{\sqrt{2}} & \sqrt{\frac{3}{2}} & -\sqrt{\frac{3}{2}} & -\frac{1}{\sqrt{2}} & ~\{r^2\}\\
            1 & -1 & -\frac{1}{\sqrt{2}} & \sqrt{\frac{3}{2}} & \sqrt{\frac{3}{2}} & \frac{1}{\sqrt{2}} & ~\{sr^2\}\\
        \end{block}
        \begin{block}{ccccccc}
            & & & & & & \\
        \end{block}
    \end{blockarray}
\end{eqnarray}
The above is a special case of (\ref{group_electric}).
Particularly, there are two condensation channels in both the top and the bottom boundaries. So the GSD is $1+1+2\times 2=6$ by (\ref{cylinder_deg2}).
The defect (confined anyon) is labeled by $S_3$ group elements, and the condensed anyon is labeled by irreducible representation of $S_3$. In idempotent completion label, these defects are 
\begin{eqnarray*} &\{e\}\equiv\{A\},~\{s\}\equiv\{D,p_1\},~\{r\}\equiv\{F,q_1\},~ \\ &\{sr\}\equiv\{D,p_2\},~\{r^2\}\equiv\{F,q_2\},~\{sr^2\}\equiv\{D,p_3\}.
\end{eqnarray*}


\subsection{Fusion of defects between $\mA_1=A\oplus B\oplus 2C/ \mA_3=A\oplus C\oplus D$}
For $\mA_3|\mA_1|\mA_3$ defects fusion, we calculate the following fusion coefficients:
\begin{eqnarray*}
    \hspace{-0.5cm}
    n_{xy}^z=\frac{\gamma_{xA}^{(3|1)} \gamma_{yA}^{(1|3)} \overline{\gamma_{zA}^{(3|3)}} } { \gamma_{ \{e\} A}^{(1|1)} } +
    \sum_{\mu,\nu=1,2}  \gamma_{x \, C^{\mu}}^{(3|1)} \left(\gamma_{\{e\}}^{(1|1)}\right)^{-1}_{C^{\mu}C^{\nu}} \gamma_{y \, C^{\nu}}^{(1|3)} \overline{\gamma_{z\,C}^{(3|3)}} 
\end{eqnarray*}
In RHS, the first part comes from shared condensed anyon $A$, in which the $M$ matrix is simply a number $\gamma_{ \{e\} A}^{(1|1)}=1/\sqrt{6}$. The second part comes from shared condensed anyon $C$, where
 the $M$ matrix is $\gamma_{\{e\}}^{(1|1)}=\matrixTwo{\sqrt{2}}{0}{0}{\sqrt{2}}$ originating from the first row of (\ref{eq:S3_AB2C_AB2C}). 
Then we have the following fusion rules:
\begin{eqnarray}
    \{A\}^{(3|1)}\otimes\{A\}^{(1|3)}&=&\{A\}^{(3|3)} \oplus \{B\}^{(3|3)} \nonumber\\
    \{F,p_1\}^{(3|1)}\otimes\{F,p_1\}^{(1|3)}&=&\{A\}^{(3|3)} \oplus \{B\}^{(3|3)} \nonumber\\
    \{F,p_2\}^{(3|1)}\otimes\{F,p_2\}^{(1|3)}&=&\{A\}^{(3|3)} \oplus \{B\}^{(3|3)}\nonumber\\
    \{A\}^{(3|1)}\otimes\{F,p_1\}^{(1|3)}&=&\{F\}^{(3|3)} \nonumber\\
    \{A\}^{(3|1)}\otimes\{F,p_2\}^{(1|3)}&=&\{F\}^{(3|3)} \nonumber\\
    \{F,p_1\}^{(3|1)}\otimes\{F,p_2\}^{(1|3)}&=&\{F\}^{(3|3)} 
\end{eqnarray}
The first fusion result is also shown in \cite{cong_defects_2017}.\\

For $\mA_1|\mA_3|\mA_1$ defects fusion, we calculate the following fusion coefficients
\begin{eqnarray*}
    n_{xy}^z=\frac{\gamma_{xA}^{(1|3)} \gamma_{yA}^{(3|1)} \overline{\gamma_{zA}^{(1|1)}} } { \gamma_{ \{e\} A}^{(3|3)} } +
    \sum_{\mu,\nu=1,2}\frac{\gamma_{x \, C^{\mu}}^{(1|3)} \gamma_{y \, C^{\nu}}^{(3|1)} \overline{\gamma_{z\,C^{\mu,\nu}}^{(1|1)}} } { \gamma_{ \{e\} C}^{(3|3)} },~
\end{eqnarray*}
and obtain the fusion rules:
\begin{eqnarray}
    \{A\}^{(1|3)}\otimes\{A\}^{(3|1)}&=&\{e\}\oplus \{sr^2\} \nonumber\\
    \{F,p_1\}^{(1|3)}\otimes\{F,p_1\}^{(3|1)}&=&\{e\}\oplus \{sr\} \nonumber\\
    \{F,p_2\}^{(1|3)}\otimes\{F,p_2\}^{(3|1)}&=&\{e\} \oplus \{s\}\nonumber\\
    \{A\}^{(1|3)}\otimes\{F,p_1\}^{(3|1)}&=&\{s\}\oplus\{r\} \nonumber\\
    \{A\}^{(3|1)}\otimes\{F,p_2\}^{(1|3)}&=&\{sr\}\oplus\{r^2\} \nonumber\\
    \{F,p_1\}^{(1|3)}\otimes\{F,p_2\}^{(3|1)}&=&\{r\}\oplus\{sr^2\} 
\end{eqnarray}

\subsection{Fusion of defects between $\mA_3=A\oplus C\oplus D / \mA_2=A \oplus B \oplus 2F $}
For $\mA_3|\mA_2|\mA_3$ fusion, the half-linking matrix (\ref{eq:S3_ACD_AB2F}) is trivial, so 
\begin{eqnarray*}
    n_{ \{A\}\{A\} }^z=\frac{ \gamma_{xA}^{(3|2)}\gamma_{yA}^{(2|3)} \overline{\gamma_{zA}^{(3|3)}} }{\gamma_{ \{e\} A}^{(2|2)}}={\tilde{\gamma}_{zA}^{(3|3)}}.
\end{eqnarray*}
The fusion rule is
\begin{eqnarray}
    \{A\} \otimes \{A\} = \{A\} \oplus \{B\} \oplus 2\{F\},
\end{eqnarray}
which can also be found in \cite{cong_defects_2017}.\\

For $\mA_2|\mA_3|\mA_2$ fusion, we have
\begin{eqnarray*}
    n_{ \{A\}\{A\} }^z=\frac{ \gamma_{xA}^{(2|3)}\gamma_{yA}^{(3|2)} \overline{\gamma_{zA}^{(2|2)}} }{\gamma_{\{A\}A}^{(3|3)}}={\tilde{\gamma}_{zA}^{(2|2)}}.
\end{eqnarray*}
The fusion rule is
\begin{eqnarray}
    \hspace{-0.8cm}
    \{A\} \otimes \{A\} = \{e\} \oplus \{s\} \oplus \{r\} \oplus \{sr\} \oplus \{r^2\} \oplus \{sr^2\}
\end{eqnarray}

\section{Non-Abelian theories $\mathcal{D} \boxtimes \bar{\mathcal{D}}$}
If the bulk theory $\mB=\mZ(\mD)$ can be factorized $\mB=\mD\boxtimes\bar{\mD}$ where $\mD$ is also a modular tensor category with modular $S$-matrix $S^{\mD}$. The anyons of $\mB$ are denoted by a pair $(a,\bar{b})$, where $a,b \in \mD$. 
Provided that both the top and bottom boundaries are described by ``diagonal condensation'', where $\mC = \{(c,\bar{c})\}$ for all $c\in \mD$, then the ``boundary defects'' are in fact the boundary excitations (simple objects) in $\mD$.

We can perform the folding/unfolding trick if $\mB$ can be factorized. Unfolding the cylinder, the doubled theory $\mB \equiv \mD \boxtimes \bar{\mD}$ unfolds into  $\mD$ over a torus. The Wilson loop $\wloop{~(x\bar{0})} \in \mB$ unfolds to $\wloop{x}\otimes \wloop{{0}}$, while the Wilson line $\wline{~~(c\bar{c})} \in \mB$ unfolds to $\wline{c}\otimes \wline{c}$.

The $S^{\mB}$ for the doubled theory is the tensor product $S^{\mB}=S^{\mD}\otimes S^{\bar{\mD}}$, so \[ S^{\mB}_{i\bar{j},k\bar{l}}=S^{\mD}_{ik} S^{\bar{\mD}}_{\bar{j}\bar{l}}=S^{\mD}_{ik}S^{\mD}_{jl}.\] The last step comes from the fact $S^{\bar{\mD}}_{\bar{j}\bar{l}}=S^{\mD}_{jl}$. 
Hence we have
\begin{eqnarray}
\frac{S^{\mB}_{x\bar{0}\,\,c\bar{c}}}{\sqrt{S^{\mB}_{0\bar{0}\,\,c\bar{c}}}}=S^{\mD}_{x\,c} = \gamma_{x\,(c\bar{c})}
\end{eqnarray}

 The $\gamma$ matrix can be identified with the $S^{\mD}$ matrix in this case. Because the cylinder basis transformation $\ket{\wloop{x}}\leftrightarrow\ket{\wline{c}}$, if viewed in the unfolded picture, is exactly the basis transformation $\ket{\wloop{x}}\leftrightarrow\ket{\wloopVtc{c}}$ on a torus, which is dictated by $S^{\mathcal{D}}$ matrix.
 In this case the Verlinde formula of boundary defects fusion is reduced to the usual Verlinde formula in topological order $\mD$.


\section{Further properties of defects fusion}
If we rewrite
\begin{eqnarray}
\gamma_{xc}=\frac{1}{\sqrt{D}}\tilde{\gamma}_{xc},
\end{eqnarray}
 where $D$ is the bulk total quantum dimension, then $\FPdim(x)=\tilde{\gamma}_{x0}$ is the quantum dimension of defect $x$. (Compare (\ref{eq:S3_ACD_AB2F},\ref{eq:S3_ACD_AFD},\ref{eq:S3_ACD_ACD},\ref{eq:S3_ACD_AB2C},\ref{eq:S3_AB2C_AB2C}) with Tab.(\ref{tab:s3_multifusion_cat}) for $D(S_3)$ defects, also see (\ref{eq:abelian_gamma_matrix}) for defects in Abelian Chern-Simons theory.)
The quantum dimension of defects is conserved during the fusion.
Taking $c=0$ in the expression 
\begin{eqnarray}
    \sum_z n_{xy}^z \gamma_{zc}^{(\mu|\rho)} = \frac{\gamma_{xc}^{(\mu|\nu)}\gamma_{yc}^{(\nu|\rho)} }{\gamma_{0c}^{(\nu|\nu)}}
\end{eqnarray}
gives 
\begin{eqnarray}
    \sum_z n_{xy}^z \FPdim(z) = \FPdim(x)\cdot \FPdim(y)
\end{eqnarray}
Note that $\gamma_{00}^{(\nu|\nu)}$ is nothing but the quantum dimension of trivial confined anyon, which comes from condensing trivial bulk anyon to boundary $\nu$. But the bulk-boundary condensation preserves quantum dimension, therefore $\gamma_{00}^{(\nu|\nu)}=1$.

\bibliographystyle{apsrev}
\bibliography{defect_Verlinde_formulaNotes}
\end{document}